\documentclass[useAMS,usenatbib]{mn2e}
\usepackage{graphicx}

\newcommand{\cgs}{$10^{-9}$ erg/s cm$^{2}$sr\,\AA\,}

\title[The 1 $\mu$m step in the cosmic background]{The 1 $\mu$m 
discontinuity in the Extragalactic Background Light spectrum: an artefact of foreground subtraction}
\author[K. Mattila]{K. Mattila\thanks{E-mail: mattila@cc.helsinki.fi}
\\
Observatory, T\"ahtitorninm\"aki, P.O. Box 14, FI-00014 University of Helsinki, Finland}
\begin{document}

\date{Accepted ; Received ; in original form }

\pagerange{\pageref{firstpage}--\pageref{lastpage}} \pubyear{2002}

\maketitle

\label{firstpage}

\begin{abstract}
Several recent papers claim the detection of a near infrared 
Extragalactic Background Light (EBL) intensity at 1.25 - 4 $\mu$m 
that exceeds the integrated light of galaxies by  factors
of $>3$. When combined with a claimed  optical 
detection of the EBL at 0.80  $\mu$m  
the EBL excess emission spectrum has a discontinuity at   $\sim 1 \mu$m. 
This discontinuity has given rise to an interpretation
in terms of ultraviolet radiation  emanating from the first
generation of massive stars at redshifts of 7 - 20 
(so called Population III stars).
The interpretation of the NIR excess emission as being of extragalactic
origin depends crucially on the model 
used in the subtraction of the Zodiacal Light, the dominant foreground
contaminant.We estimate the Zodiacal Light at 0.80  $\mu$m   
using on the one hand the measurement by \citet{b2},
with corrections for some omitted effects of atmospheric scattering 
and calibration, and on the other hand the model of \citet{k1}.
There is in neither case any evidence for a step in the EBL at $\sim 1 \mu$m.
We emphasize that in order to avoid systematic effects it is essential 
to use the same  Zodiacal Light model \citep{k1} 
for both the NIR (1.25 - 4 $\mu$m) and optical (0.80  $\mu$m) data.
We emphasize, however, that our analysis does not allow a statement
on the {\em overall level} of the NIR EBL.  
The contribution of the Diffuse Galactic Light to the
``EBL excess'' emission is estimated. It is found to be significant at  3 - 4 $\mu$m
and should be carefully evaluated in future measurements which 
aim at detecting an EBL signal at the level of $\sim$10 nW\, m$^{-2}$ sr$^{-1}$,
i.e. at the level of the integrated light of (known) galaxies.

\end{abstract}


\section{Introduction}

The Extragalactic Background Light (EBL) in the ultraviolet, optical, 
and infrared wavelength regions can potentially provide a measure 
for the total luminous matter in the Universe since the epoch of
star and galaxy formation; for a review see \citet{lo1}.

Detection of a near infrared Extragalactic Background Light (EBL) component,
in excess of the integrated light from known galaxies, has
recently been announced by several research groups using the {\em COBE}/DIRBE
\citep[for a review see ][]{dw05} and the Japanese {\em IRTS} instruments \citep{m1}. 
A ``first detection'' of the optical EBL
at 0.30, 0.55, and 0.80 $\mu$m has been claimed by  \citet{b1,b2}
(hereafter referred to as BFM02a,b).
Combining the claimed detections in these two wavelength bands,
i.e. at 1.25 - 4  $\mu$m and 0.8 $\mu$m, there appears to be
an indication of a discontinuity at $\sim 1 \mu$m,
the EBL brightness rising by a factor of  $\sim 4$
from the short-wavelength to the long-wavelength side of this step 
\citep[see e.g.][]{m1}.
The potential importance of this spectral discontinuity stems from
its suggested interpretation as signature of the 
first generation of stars (s.c. Population III stars) with their
rest-frame emission at $\lambda > 0.912  \mu$m redshifted 
by a factor of $1+z \approx 8 -20$ to the near infrared 
\citep{santos,salvaterra, cooray, madau,dw05}. 

The reality of the NIR EBL excess at 1 -- 4~$\mu$m  and the 1 $\mu$m spectral step
have been contested both on observational \citep[see e.g. ][]{ah,dw05}
and on theoretical grounds \citep[see e.g. ][]{madau}.
The present author \citep{m2} has argued that, because of insufficient
or incorrect  atmospheric scattering corrections and calibration,
the claim by BFM02 for the detection of the EBL at  0.30, 0.55, and 0.80 $\mu$m
was premature. 

In the present paper we pose the question whether the BFM02 measurement 
can be used to set a sufficiently low {\sl upper limit} to the EBL at  
0.80 $\mu$m which, 
in combination  with the claimed NIR EBL values at 1.25 - 4  $\mu$m,
would justify the claim for a  step near $\sim1 \mu$m in the 
EBL spectrum. Because of the large intensity of the 
Zodiacal Light in comparison with the EBL its
accurate subtraction is crucial for any statement on the EBL and 
its spectral shape. Therefore,
we examine in detail the underlying ZL determinations and models used 
in the above mentioned works. We will also examine the contribution
of the Diffuse Galactic Light (DGL) to the observed 0.8 -- 4~$\mu$m
sky brightness.

\section{Zodiacal Light estimation}

The basic formula used in the above mentioned papers for extracting 
the NIR or optical EBL is the following:
\begin{equation}
 I_{EBL} = I_{tot} - I_{ZL} - I_{ISL} - I_{DGL} 
\end{equation}

where  $I_{tot}$ is the total sky brightness as 
measured from space, $I_{ZL}$ the Zodiacal Light (ZL) as
estimated using dedicated observations or a model of the interplanetary 
dust cloud,  $I_{ISL}$
is the Integrated Starlight (ISL) from unresolved stars evaluated from 
star count models, and  $I_{DGL}$ is the Diffuse Galactic Light (DGL).

The first two terms are large, $\sim$300 nW\, m$^{-2}$ sr$^{-1}$ at 
1 - 2 $\mu$m, as compared to the EBL which according to e.g. 
\citet{m1} and \citet{c1} is 
 $\sim$50-70 nW\, m$^{-2}$ sr$^{-1}$ , 
but has conventionally been considered to be much smaller, 
$\sim$10 nW\, m$^{-2}$ sr$^{-1}$,
and mainly due to the integrated light of galaxies only.
While the ISL term is smaller, it is still several times larger
than the EBL signal  due to known galaxies. 
Thus, the demands for the accuracy of the  $I_{tot}$ and $I_{ZL}$ 
determinations are extremely high. 

\subsection{ZL estimated from the observations of \citet{b1,b2}}

In the case of the claimed EBL detections at 0.30, 0.55, and 0.80 $\mu$m
by BFM02 the  $I_{ZL}$ term was derived from {ground based
observations} of the depths of Fraunhofer lines between 0.39 and 0.51 $\mu$m.
Large uncertainties are caused to ground based night sky photometry
by the corrections needed for the atmospheric scattered light and
extinction. Also, the method necessitates to calibrate
the {\em extended source} responses of 
two different telescopes, a ground based and a space borne one,
to the same {\em absolute} scaling. 

In a re-discussion of the BFM02 ZL observations 
\citet{m2} showed that corrections had to be applied  
because of the following errors or omissions:  
(1) incorrect aerosol albedo;
(2) omission of ground reflectance; (3) omission of  DGL as source
of atmospheric scattering; (4) incorrect aperture-correction factor.
All  these corrections were negative in sign and thus had the effect of decreasing
the ZL value. Their combined effect to  
$I_{ZL}$(0.80 $\mu$m) amounted to -9.4 to -13.1 $10^{-9}$ erg/s cm$^{2}$sr\,\AA\,,
see Table 1 of \citet{m2}. This correction is ca. 10 times larger than the accuracy of 
$\pm 1\, 10^{-9}$ erg/s cm$^{2}$sr claimed by BFM02a for their
ZL value and ca. 4-6 times as large as their claimed
EBL value of  2.2\,$10^{-9}$ erg/s\,cm$^{2}$\,sr\,\AA\,  at 0.80\,$\mu$m.
Each one of the corrections (1)--(4) has large uncertainties caused
by the ill-defined atmospheric, ground reflectance and telescope PSF
properties. We estimate that the resulting correction has
an uncertainty of $\sim\pm$50\% or   
$\sim\pm$5 $10^{-9}$ erg/s cm$^{2}$sr\,\AA\,. 
In an Erratum paper \citet{b5} discussed a number of errors and 
omissions in BFM02.
While some of the errors noted by  \citet{m2} were addressed, they did not yet
address the four major corrections (1) - (4) as discussed above. 
The original ZL estimate of BFM02b is given in line 2 of Table 1
and the \citep{m2} estimate, including these corrections (1) - (4), 
in line 4 of Table 1.

Another point of concern is how BFM02 handled the {\em combination} of 
the individual systematic errors: this was done by assuming a flat probability 
distribution for each contributing source of error. This method produces 
combined systematic errors which are by factor 2 -- 3 smaller than the 
conventionally determined errors obtained by adding in quadrature 
the individual systematic errors. One specific systematic error,
important for the present purpose of estimating $I_{ZL}$ at 0.80\,$\mu$m,
is caused by the reddening of the ZL relative to the Solar colour. 
BFM02a estimate the reddening correction factor per 1000 \AA\, to be
$C(\lambda) = 1.044$ with a systematic error of $[-0.014, +0.006]$.
Extrapolating  $I_{ZL}$ from the reference wavelength of 
 0.465\,$\mu$m to  $0.80\,\mu$m 
the systematic error caused by the uncertainty in the ZL reddening
correction thus becomes $[-4.1\%,+1.75\%]$. Combining this {\em in quadrature}
with the other systematic errors the total systematic uncertainty
of  $I_{ZL}$(0.8$\mu$m) becomes  $[-4.4\%,+2.4\%]$ 
or $[-3.0,+1.6]$ \cgs (line 3 of Table 1).

\begin{table*}
 \centering
 \begin{minipage}{185mm}
  \caption{Estimation of the Zodiacal Light and Extragalactic Background Light for the BFM02 Field at 
           0.8$\,\mu$m using the ground-based ZL observations of \citet{b2} (BFM02), with corrections according to \citet{m2} (M03). 
Systematic errors are given in brackets. }
  \begin{tabular}{@{}lllll@{}}
  \hline
   & Quantity      &   Unit         & Value & Comment\\ 
 \hline
1& $I_{tot}^{\lambda}$(HST), 8000 \AA\, & $10^{-9}$ erg/s cm$^{2}$sr \AA\, & $72.4\pm0.14, [\pm$1.0] &BFM02, with ``1-$\sigma$'' systematic error\\
\hline
2&$I_{ZL}^{\lambda}$(0.80$\,\mu$m)  &  $10^{-9}$ erg/s cm$^{2}$ sr \AA\,& $69.4\pm0.4$,$[-0.9,+0.8]$ & BFM02, with ``1-$\sigma$'' systematic error \\
3&$I_{ZL}^{\lambda}$(0.80$\,\mu$m)  &  $10^{-9}$ erg/s cm$^{2}$ sr \AA\,& $69.4\pm0.4$,$[-3.0,+1.6]$ & BFM02, but with conservative systematic error \\
4&$I_{ZL}^{\lambda}$(0.80$\,\mu$m)  &  $10^{-9}$ erg/s cm$^{2}$ sr \AA\,& 56.3-60.0  & BFM02 as corrected by M03 \\
\hline
5&$I_{EBL+DGL}^{\lambda}$(0.80$\,\mu$m) &  $10^{-9}$ erg/s cm$^{2}$ sr \AA\,& $3.0\pm0.4,[\pm0.9]$ &
 BFM02, with 1-$\sigma$ statistical and  \\
 &                                     &                     &     &   ``1-$\sigma$'' systematic errors \\
6&$I_{EBL+DGL}^{\lambda}$(0.80$\,\mu$m) &  $10^{-9}$ erg/s cm$^{2}$ sr \AA\,& $3.0\pm0.8, [-1.9,+3.2]$ 
& BFM02, but with 2-$\sigma$ statistical \\
   & & & & and  conservative systematic errors  \\
7&$I_{EBL+DGL}^{\lambda}$(0.80$\,\mu$m) &  $10^{-9}$ erg/s cm$^{2}$ sr \AA\,&$<(12.4-16.1)$   
& BFM02 with corrected ZL from M03\\
8&$\lambda\cdot I_{EBL+DGL}^{\lambda}$(0.80$\,\mu$m) & nW/m$^{2}$ sr &$<(99-129)$
& BFM02 with corrected ZL from M03\\
\hline
\end{tabular}
\end{minipage}
\end{table*}

\begin{table*}
 \centering
 \begin{minipage}{185mm}
  \caption{Estimation of the Zodiacal Light and Extragalactic Background Light for the BFM02 Field at 
           0.8$\,\mu$m using the \citet{k1} (K98) Zodiacal Light model. Systematic errors 
are given in brackets.}
  \begin{tabular}{@{}lllll@{}}
  \hline
   & Quantity      &   Unit         & Value & Comment\\ 
 \hline
1& $I_{tot}^{\lambda}$(HST), 8000 \AA\, & $10^{-9}$ erg/s cm$^{2}$sr \AA\, & $72.4\pm0.14, [\pm$1.0] &BFM02, with ``1-$\sigma$'' systematic error\\
\hline
2&$I_{ZL}^{\nu}$(model), J band  & MJy sr$^{-1}$    & $0.151,[\pm0.0062]$ & K98 model value from DSZA\\
3&Colour correction factor K     &                  & 0.985        & Assuming 6000 K blackbody\\
4&$I_{ZL}^{\nu}$(1.25$\,\mu$m)    &  MJy sr$^{-1}$ & $0.153,[\pm0.0063]$ & Monochromatic intensity\\
5&$I_{ZL}^{\lambda}$(1.25$\,\mu$m)  & $10^{-9}$ erg/s cm$^{2}$ sr \AA\,  & $29.4,[\pm1.15]$ & Monochromatic intensity\\
6&$F_{\sun}^{\lambda}$(1.25$\,\mu$m) & erg/s cm$^{2}$ \AA\, &  44.6& Monochromatic Solar flux\\
7&$F_{\sun}^{\lambda}$(0.80$\,\mu$m) & erg/s cm$^{2}$ \AA\, & 112.3& Solar flux through WFPC f814\\
8&$F_{\sun}^{\lambda}$(0.80$\,\mu$m)/$F_{\sun}^{\lambda}$(1.25$\,\mu$m) &     &  2.52 & Solar SED\\
9&$I_{ZL}^{\lambda}$(0.80$\,\mu$m)/$I_{ZL}^{\lambda}$(1.25$\,\mu$m)    &      & 2.26$[\pm0.05]$ & Reddened Solar SED\\
10&$I_{ZL}^{\lambda}$(0.80$\,\mu$m)  &  $10^{-9}$ erg/s cm$^{2}$ sr \AA\,& $66.4[\pm2.6],[\pm1.5]$& ZL resulting from the K98 model\\
\hline
11&$I_{EBL+DGL}^{\lambda}$(0.80$\,\mu$m) &  $10^{-9}$ erg/s cm$^{2}$ sr \AA\,& $6.0\pm0.1,[\pm3.2]$ &
 $I_{tot}^{\lambda}$(HST) {\sl minus} $I_{ZL}$(K98)\\
12&$\lambda\cdot I_{EBL+DGL}^{\lambda}$(0.80$\,\mu$m) & nW/m$^{2}$ sr & $48.0\pm25.6$ & 
 $\lambda\cdot I_{tot}^{\lambda}$(HST) {\sl minus} $\lambda\cdot I_{ZL}$(K98)\\ 
\hline
\end{tabular}
\end{minipage}
\end{table*}

\subsection{ZL estimated using the \citet{k1} model}

The Zodiacal Light model which has been adopted in
many of the NIR EBL studies, e.g. \citet{m1} and \citet{c1},
has been constructed by \citet{k1}. This model
uses the full {\em COBE}/DIRBE data set at 10 wavelengths between
1.25 and 240  $\mu$m and relies essentially on the seasonal
variations of the ZL. The IPD emission dominates the sky brightness
at  high galactic latitude areas and mid IR wavelengths,  
$\lambda \approx 10 - 60 \mu m$, a situation which helps to 
constrain the model parameters at these wavelengths. 

In order to investigate whether the different approaches
used for the ZL estimation at $\lambda \approx 1 - 4 \mu m$ (modelling)
and at  $\lambda = 0.80 \mu m$ (ground based observations, BFM02b) could 
be the reason for the $\sim$1$\mu$m EBL spectrum step   
we have applied the \citet{k1} ZL model to the $I_{tot}$ 
observation of BFM02a at  $0.80 \mu m$. To do this we have extracted from the DIRBE 
Sky and Zodi Atlas (DSZA) the ZL value at J band ($1.25 \mu$m)
for the specific sky position ($\lambda$ - $\lambda_{\sun}$, $\beta$) 
and day of the year of the {\em HST} measurement of $I_{tot}$ by BFM02a.
This ZL value, in units of MJysr$^{-1}$, is given in line 2 of Table~2.
The systematic error estimate given in brackets corresponds to 
the error of 15 nW/m$^{2}$ sr$^{-1}$ as given  in Table 7 of \citet{k1}.
To transform the tabulated DSZA value into a monochromatic intensity
at 1.25 $\mu$m a small colour correction is applied assuming 
for the ZL a solar temperature blackbody spectrum.

To estimate the ZL intensity at 0.80 $\mu$m 
we adopt the Solar SED
as given by \citet{c2}, convolve it with the HST WFPC2 filter F814W
transmission curve using the SYNPHOT/calcphot 
procedure\footnote{http://www.stsci.edu/resources/software\_hardware/stsdas/synphot},
and account for the  ZL reddening relative to the Solar SED
by applying a correction factor according to Eq.(22) of \citet{l1}.  
The reddening correction 1.25\,$\mu m$ $\rightarrow$  0.80\,$\mu m$ 
of 1.116 is estimated to have a 
{\sl systematic} error of $\pm0.023$, i.e. $\pm$20\%.
The resulting ZL brightness of $66.4[\pm2.6],[\pm1.5]$ at 0.80 $\mu$m 
is given in line 10 of Table 2. The first error in brackets corresponds 
to the \citet{k1} model uncertainties, the second one to the 20\% error 
adopted for the ZL reddening correction.

\subsection{The residual signal, EBL + DGL, after subtraction of the ZL}

BFM02a give for their {\sl HST} 
measurement of  $I_{tot}$ at 0.80\, $\mu$m  the value 72.2$\pm$0.14 $[\pm1.0]$ \cgs
(line 1 of Table 1 and Table 2).  
Their method of treating systematic errors
leads to the paradoxical result (line 5 of Table 1) that subtracting
the value of $I_{ZL}$(0.8$\mu$m) = 69.4 $\pm0.4 [-0.9, +0.8]$  
$10^{-9}$ erg/s cm$^{2}$sr\,\AA\,
leaves as difference the value $I_{EBL+DGL}$(0.8$\mu$m) = 3.0  $\pm0.4 [\pm0.9]$  
$10^{-9}$ erg/s cm$^{2}$sr\,\AA\,,
i.e. the systematic error has {\em decreased} from its level for  
$ I_{tot}$(0.8$\mu$m) alone. We give in line 6 of Table 1 a more conservative error 
estimate with $2\sigma$ statistical and quadratically added systematic errors.
The systematic errors of $\sim\pm$5 \cgs caused by the four corrections discussed
by \citet{m2} and in Sect. 2.1 above have not been included here. Instead, lines 7 
and 8 give the estimated upper limits of  $I_{EBL+DGL}^{\lambda}$ and  
$\lambda\cdot I_{EBL+DGL}^{\lambda}$, reflecting the conclusion of \citet{m2}
that the claim for a detection of the EBL by BFM02 was premature.

Subtraction of the \citet{k1} $I_{ZL}$ estimate from $I_{tot}$  
results in the values of $I_{EBL+DGL}^{\lambda}$ and  
$\lambda\cdot I_{EBL+DGL}^{\lambda}$ as given in lines 11 and 12 of Table 2.
The systematic errors of  $I_{tot}$ and  $I_{ZL}$ have been added
in quadrature.

Fig.~1 shows the  $\lambda\cdot I_{EBL+DGL}^{\lambda}$ values and upper limits.
They include the {\em IRTS} Near-Infrared Spectrometer (NIRS) values of \citet{m1}
between 1.4 - 4 $\mu$m and the {\em COBE}/DIRBE values at 1.25, 2.2, 3.5, and 4.9  $\mu$m
as compiled in \citet{dw05} and \citet{h01}.
Our two estimates at 0.8 $\mu$m are shown as an upper limit (box), based on the 
BFM02b ZL estimate as revised by \citet{m2},
and as an open circle with error bars, based on the ZL estimate from the \citet{k1}   
model.

\begin{figure}
\includegraphics[width = 7.8cm]{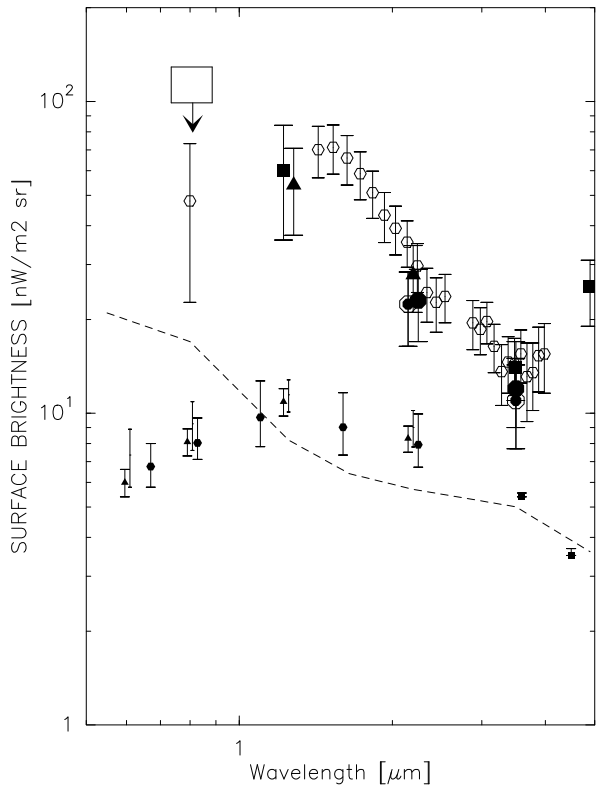}
 \caption{Observed values and upper limits for the Extragalactic Background 
Light in the red and near infrared $\lambda$ = 0.8-5  $\,\mu$m. The two 
estimates at  0.8$\,\mu$m derived in this paper are indicated by an open circle
and error bars \citep[ZL model of][]{k1} and a rectangular box 
\citep[BFM02 ZL observations with corrections according to ][]{m2}.
The open circles with error bars at 1.4-4 $\mu$m  are the  {\em IRTS} Near-Infrared Spectrometer 
(NIRS) values of \citet{m1}. The {\em COBE}/DIRBE values at 1.25, 2.2, 3.5, 
and 4.9  $\mu$m are indicated by big solid symbols: squares  \citep{dw98, ar03},
 triangles \citep{c1},  octagons  \citep{w1}, and encircled octagons \citep{g1}.
The integrated light from known galaxies
is shown by small solid symbols: hexagons \citep{mapo}, triangles \citep{totani} and 
squares \citep{fa}. The estimated lower and upper limit for the {\it total} light
of galaxies is shown by  bars at 0.61, 0.81, 1.25, and 2.2 $\mu$m \citep{totani}.
Our estimate for the Diffuse Galactic Light,  $I_{DGL}$(sca+em), is shown as a
dashed line according to column 11 of Table 3}
\end{figure}

\begin{table*}
 \centering
 \begin{minipage}{185mm}
  \caption{Estimation of the diffuse Galactic surface brightness due to
scattered starlight,  $I_{DGL}$(sca),  and  non-equilibrium emission or fluorescence by dust,  $I_{DGL}$(em). }
  \begin{tabular}{@{}lllllllllll@{}}
  \hline
Field & $l, b$ & $N$(H)&$A_V$& $\lambda$ & $\lambda I^{\lambda}_{ISRF}$ & $A_{\lambda}$ & $\lambda I^{\lambda}_{ISRF}\, A_{\lambda}$ &  $\lambda I^{\lambda}_{DGL}$(sca)  & $\lambda I^{\lambda}_{DGL}$(em) &  $\lambda I^{\lambda}_{DGL}$(sca+em)\\
      & deg & 10$^{20}$\,cm$^{-2}$ &mag&  $\mu$m    & nW/m$^{2}$sr & mag  & \multicolumn{4}{c}{nW/m$^{2}$sr}   \\
 (1)  & (2) & (3) & (4) & (5) & (6) & (7) & (8) & (9) & (10) & (11) \\ 
\hline
BFM02 & 206.6, -59.8 &1.5& 0.080& 0.55 & 660\footnote{Mattila 1980}&0.080&53& 21\footnote{derived from $I_{DGL}(sca)/I(100\mu$m) in L~1642 \citep{lms}} & -- & 21  \\
  &  & &                        & 0.80 & 867$^a$(900)                   &0.048&42& 17 (8.0\footnote{effective value to be applied to $I_{EBL+DGL}$ of BFM02})&-- & 17   \\
\hline 
Matsumoto    &   80-90, 42-48 &1.3& 0.069& 1.25 & 1060\footnote{\citet{lm}}&  0.019   &  20  & 7.9                 & 0.3   & 8.2    \\
et al. (2005)&              &  &   &       1.65 & 1110$^d$&0.012&13 & 5.2                 & 1.2   & 6.4     \\
             &              &  &   &       2.2  & 655$^d$ &  0.0075  &  4.9 & 1.9                 & 3.8   & 5.7    \\
             &              &  &   &       3.5  &  &    &   &                 &    & 5.0       \\
             &              &  &   &       4.9  &  &   &     &                  &    & 3.6       \\           
\hline
\end{tabular}
\end{minipage}
\end{table*}

\section{The contribution by the Diffuse Galactic Light}

We have so far presented in Tables 1 and 2 and Fig. 1 the combined contribution 
of the EBL and the DGL. Since \citet{m1} do not make any statement on the DGL
they appear to have included it into their nominal ``EBL'' values.
For the  {\em COBE}/DIRBE based values at 1.25, 2.2, 3.5, and 4.9
shown in Fig.~1 the DGL has been considered to different degrees.
The following discussion in this section is concerned with the DGL
contribution to the  \citet{m1} 1.4 -- 4$\mu$m and the BFM 0.80$\mu$m values.

BFM02a used a general model of the ISRF and dust distribution to estimate
the DGL (scattered starlight) contribution in their target field
at  0.80 $\mu$m: $I_{DGL} = 0.8$ \cgs. \citet{m2} revisited the model
estimate and argued that the dust column density toward the 
target field had been underestimated by a factor of 
$\sim3$ and estimated the DGL intensity at  
0.8$\mu$m to be $I_{DGL} = 2.3$ \cgs corresponding to 18  nW/m$^{2}$sr.
However, after including a correction caused by the
fact that the DGL has a Fraunhofer line spectrum resembling 
 the ZL spectrum \citet{m2} found that an  {\em effective}  
$I_{DGL}$ correction of $\sim 1$ \cgs or 8 nW/m$^{2}$sr
shall be subtracted from the observed $I_{EBL+DGL}$ value of BFM02. 

At optical and NIR wavelengths up to 1.5 $\mu$m  the diffuse Galactic surface 
brightness is primarily due to Galactic starlight scattering off interstellar dust grains. 
At longer wavelengths ($\sim\,2-5 \mu$m), because of decreasing optical
depth of dust ($A_{\lambda} \la 0.15\,A_V$), the scattered light contribution
rapidly decreases. However, recent {\em COBE}/DIRBE and {\em Spitzer} observations 
have shown that 
there is at NIR wavelengths, $\lambda >  2 \mu$m, a substantial diffuse  
Galactic surface brightness  component \citep{be,a1,da03,bo}
associated with the widely distributed low density dust medium.
Such a NIR continuum emission had been previously detected in reflection nebulae 
by \citet{s1}. It can be due e.g. to non-equilibrium emission
by very small transiently heated grains or to fluorescence emissions
by PAHs or very small grains; for a discussion see \citet{bo}.

We will estimate the diffuse Galactic emission 
at the different wavelengths and for the two regions of sky 
targeted in BFM02 and \citet{m1} observations. 
The line-of-sight dust column density can be estimated using the
21-cm hydrogen column density map of \citet{ka}. 
For this map the best state-of-the-art 
stray radiation corrections have been applied, important for
the high latitude fields with weak H{\sc I} emission.
The H{\sc I} column densities for the two fields are 1.5 and 1.3
 10$^{20}$\,cm$^{-2}$ as given in 
column(3) of Table 3. Using the standard $N$(H)/$E(B-V)$ ratio \citep{boh}
and $R_V = 3.1$ they correspond to the line-of-sight
extinctions of $A_V$ = 0.080 and 0.069 mag, respectively.
We assume that the scattering properties of the dust grains, 
i.e. albedo $a$ and forward scattering parameter $g$, are approximately
constant between 0.55 and 2.2 $\mu$m \citep[for justification see][]{lm}.
Then, the scattered light intensity at different wavelengths is,
for an optically thin line of sight towards high galactic latitudes,
in a first approximation proportional to the product 
${I({\rm ISRF})}\times A_{\lambda}$. We give in columns (6)-(8)
of Table 3 our estimates for ${I({\rm ISRF})}$ and  $A_{\lambda}$.
The wavelength dependence of  $A_{\lambda}$ corresponds to $R_V = 3.1$
\citep{mathis}. For  ${I({\rm ISRF})}$ we have adopted at $1.25-2.2 \mu$m
the mean sky brightness values as obtained from 
 the COBE/DIRBE Zodi Subtracted Mission Averaged (ZSMA) maps; see \citet{lm}. 

The scattered light intensity, $I_{DGL}$(sca), has been measured
in the translucent high latitude ($b = -36.7$ deg) cloud L~1642
as a function of optical thickness for five wavelengths between
$\lambda$ = 0.35 - 0.55  $\mu$m \citep{lms,m90}.
At small optical depths we obtain from these observations
at 0.55 $\mu$m the relation  $I_{DGL}$(sca)/$I(100 \mu$m) = 3.0 \cgs/MJy~sr$^{-1}$ 
which for $I(100 \mu$m)$/N$(H{\sc I}) = 0.85 MJy~sr$^{-1}$/10$^{20}$\,cm$^{-2}$
\citep{bou96} corresponds to  $I_{DGL}$(sca)/N$(H{\sc I})$ = 
2.55 \cgs/10$^{20}$\,cm$^{-2}$. Thus, for the BFM02 field 
we obtain at  0.55  $\mu$m the 
estimate $\lambda I^{\lambda}_{DGL}$(sca) = 21  nW/m$^{2}$sr as given in column (9) of Table 3.
The  $\lambda I^{\lambda}_{DGL}$(sca) values for the other wavelengths in Table 3 are then
derived by scaling  this $\lambda I^{\lambda}_{DGL}$(sca) at  0.55  $\mu$m with the
 $\lambda I^{\lambda}_{ISRF} \times A_{\lambda}$ values given in column (8).
The value thus obtained for the BFM02 field at 0.80 $\mu$m is
17 nW/m$^{2}$sr which very closely corresponds to the model
estimate of BFM02a as corrected by \citet{m2}.
These optical DGL estimates are in general terms also confirmed by the surface
photometry of L~1780 \citep{m79} which stretched over the larger wavelength range
of 0.35 - 0.755 $\mu$m but lacked the smallest optical depth range
important for the present purpose.
It can be seen from Table 3 column (9) that $\lambda I^{\lambda}_{DGL}$(sca)
drops rapidly at $\lambda > 1.5 \mu$m.

We use the results of \citet{bo} 
to estimate the  $\lambda I^{\lambda}_{DGL}$ values at 1.25 - 4.9 $\mu$m  for 
the \citet{m1} field.
In their Fig. 9 they give the diffuse Galactic brightness at 
2 - 14 $\mu$m for  $N$(H) =  10$^{21}$\,cm$^{-2}$. 
From this figure we read the  
$I_{DGL}$(tot) = $I_{DGL}$(sca) +  $I_{DGL}$(emission) values
for 2.2, 3.5, and 4.9  $\mu$m and they are given in column (11)
of table 3. In order to
estimate $I_{DGL}$(emission) at 1.25 and 1.65  $\mu$m we use the
 \citet{bo} model for the NIR continuum which they represent
with a grey body of colour temperature 1100$\pm$300 K.
The values thus obtained for  $\lambda I^{\lambda}_{DGL}$(emission) 
are given in column (10) of Table 3. At these two wavelengths  $I_{DGL}$(sca)
already dominates over  $I_{DGL}$(emission).
It can be seen from Table 3 that the total  $\lambda I^{\lambda}_{DGL}$ in units of  nW/m$^{2}$sr  
varies by a factor of $\sim$2 over the NIR wavelength range, $\lambda$ = 1.25 - 4.9 $\mu$m.
At 0.80  $\mu$m it is $\sim 2$ times as large as at  1.25 $\mu$m. However,  after 
including the correction caused by the similarity of the DGL and ZL Fraunhofer line spectra
its  {\em effective} value, to be subtracted from
the BFM02  $\lambda I^{\lambda}_{EBL+DGL}$ value at 0.80 $\mu$m, becomes 8 nW/m$^{2}$sr,
i.e. very similar to the  value at 1.25 $\mu$m.

The DGL contributions are, in general, substantially smaller than the  
 $\lambda I^{\lambda}_{EBL+DGL}$
values of BFM02 and \citet{m1} shown in Fig. 1. The values are, however, 
closely similar to the
integrated intensities derived from galaxy counts. Therefore, if the 
true NIR EBL surface brightness turns out to be of this order of magnitude
the DGL contribution cannot be neglected.
Rather, it will be the ultimate obstacle, even for space borne 
measurements from outside the Zodiacal cloud, which must be accurately
known and subtracted before arriving at a credible measurement of the EBL.

\section{Discussion}

\subsection{Is there an $\sim$1$\,\mu$m step in the EBL spectrum?}

The interpretation of the NIR EBL excess as signature 
of Population III stars hinges on the reality of the claimed large 
discontinuity between 0.8 and 1.2-1.4 $\mu$m.
The values (upper limits) derived in Section 2 and shown in Fig. 1 
for  $\lambda I^{\lambda}_{EBL+DGL}$ at 0.8   $\mu$m  are significantly
larger than the original BFM02a value of $3.0\,10^{-9}$ erg/s cm$^{2}$ sr \AA\,
or 24  nW/m$^{2}$sr on which this conjecture was based.

The large range of the estimates at  0.80$\mu$m reflects directly the differences and 
uncertainties of the values of $I_{ZL}$(0.80$\,\mu$m) as derived 
from the \citet{k1} model on the one, and  from the BFM02b ground based observations
on the other side.
In both cases the ZL value has been extrapolated from another
wavelength,  1.25$\,\mu$m in the former and 0.465$\,\mu$m in the latter case,
using a reddened Solar SED to transform from Solar to ZL colour.

When assessing the reality of the $\sim$1$\,\mu$m step in the EBL spectrum
it is obvious that the subtraction of the foreground ZL has to be done
using consistent ZL values, both on the short and on the long wavelength side 
of the step. With this in mind we have applied the \citet{k1} model
and it can be seen from Fig. 1 that there is, within the error limits, no
evidence for the step.
Errors in the basic assumptions or parameter values of the \citet{k1} model would 
influence the values at  $0.80\,\mu m$ and at $> 1\mu$m in a similar way
and are not expected to artificially create or destroy a large spectral step,
if existent, between these wavelengths.

When applying the BFM02b measurement of $I_{ZL}$  
an extrapolation of their measured ZL value from  0.465$\,\mu$m to 0.80$\,\mu m$ is 
needed. This introduces an uncertainty of similar size as the
extrapolation of the \citet{k1} model from 0.80 to 1.25 $\mu$m.
Given the corrections needed for the BFM02b ZL measurement and adopting 
conservative systematic error estimates as discussed in Sect. 2.1 above,
there is no evidence for a downward step in the EBL spectrum 
from 1.25 to 0.80 $\mu$m.

The Diffuse Galactic Light correction as discussed in Sect. 3 is of similar 
strength both below and above  1$\,\mu$m  and it does not, therefore, 
contribute to the possible discontinuity at  $\sim$1$\,\mu$m.

The author wishes to emphasise that he does not advocate the (very) large
$I_{EBL+DGL}$ value at 0.80$\mu$m, apparently resulting from the discussion 
in this paper, as anything else but an upper limit. Rather, also the value derived using 
the \citet{k1} ZL model, which appears to be different from zero, should be considered 
only as an upper limit. 

\subsection{Remarks on the near infrared EBL measurements}

 The analysis presented in this paper does not allow  conclusions
concerning the {\em overall level} of the NIR EBL at 1 - 4\,$\mu$m.
Given the systematic uncertainties of the ZL subtraction
a direct photometric detection of the NIR EBL appears very difficult.

The \citet{k1} model appears to give a good representation
of the ZL thermal emission at mid and far IR wavelengths where the
ZL dominates the sky brightness. However, at the near IR wavelengths, 
$1 \mu$m $< \lambda < 5\mu$m, the Zodiacal component
is weaker and its modelling relies on the mid IR results.
As stated by \citet{k1} their ZL model is
not unique but the predicted values depend on the assumed
models for the IPD cloud geometry and dust parameters.
Furthermore, the NIR ZL is caused by scattering instead 
of thermal emission which means that dust parameters different 
from those at the mid IR wavelengths have to be used.

An alternative ZL model has been presented by \citet{w98} and has been
applied in the \citet{wr00}, \citet{w1}, and \citet{g1} analyses
of the DIRBE data. The resulting EBL values are by $\sim$33, 8, and
4  nW/m$^{2}$sr smaller at the J, K, and L bands, respectively, as 
compared to the results based on the \citet{k1} model. This does not, however, 
substantially change the dilemma with an unexpectedly large NIR EBL value. 

\citet{dw05} have presented arguments suggesting that the
\citet{m1} EBL excess emission at  $\lambda = 1.4 - 4\,\mu m$ may be
due to insufficient ZL subtraction. This possibility is especially
suggested by the practically identical SEDs of the ZL and the ``EBL excess''
over the whole wavelength range of the observations,
see Fig. 6 of \citet{dw05}. 

The ``EBL excess'' emission as derived from the {\em IRTS} data by 
\citet{m1} contains the contribution by the Diffuse Galactic emission. 
The DGL has been considered in several of the analyses of the 
{\em COBE}/DIRBE data \citep[see e.g.][]{a1, ar03} but its separation
from the much larger component due to the Integrated Starlight has been
problematic because of the large field of view of DIRBE. The recent
{\em Spitzer} observations \citep{bo} have offered a better estimate
for the 2-5 $\mu$m DGL and they suggest that it is substantially 
larger than the values previously found by \citet{a1}. Although the DGL level
remains insignificant over the wavelength range  1.25-2.5 $\mu$m
this is no longer the case at 3.5 $\mu$m. At this wavelength
the sum of the integrated light from resolved galaxies \citep{fa} 
and the DGL (Table 3) amounts to $\sim$10.4 nW/m$^{2}$sr which is
within its estimated error limits of $\sim\pm2$  nW/m$^{2}$sr fully 
compatible with the \citet{m1} EBL value of $14.4\pm 3$  nW/m$^{2}$sr.
In addition, similarly as at the shorter wavelengths \citep{totani}, 
some contribution from the unresolved galaxies should be
added \citep{kash,savage} 
bringing the two estimates even closer to each other. 

 Recently, an upper limit to the EBL flux around  1 - 2$\mu$m 
has been announced by \citet{ah} using {\em HESS} observations of intergalactic absorption 
of $\gamma$-ray emission of blazars: $I_{\rm EBL} \leq (14\pm4)$ nW/m$^{2}$sr.  
Thus, these {\em HESS} observations would suggest that more than two thirds
of the EBL in the NIR band is resolved into individually detected galaxies.
The  {\em HESS} result depends, however, on the adopted TeV spectra of the
blazars. Therefore, direct photometric observations of the optical/NIR
EBL are still urgently needed.

\section{Conclusions}

We have scrutinised the observational evidence for the claimed discontinuity
at  $\sim$1$\mu$m in the spectrum of the Extragalactic Background Light.
We have also estimated the contribution of the Diffuse Galactic emission
to the ``EBL excess''. Our conclusions are the following:\\
(1) The reality of the  $\sim$1$\mu$m step hinges on the 
claimed detection of the EBL by BFM02 at 0.80 $\mu$m. We find that
after applying  corrections to the analysis of their 
Zodiacal Light measurement only an upper limit to the EBL can
be set which does not imply any spectral step between  0.80 and 
1.25 $\mu$m.\\
(2) An alternative estimate of the Zodiacal Light at  0.80 $\mu$m
is obtained using the  \citet{k1} model. Again, the upper limit
set to the EBL at 0.80 $\mu$m does not warrant any spectral step between  0.8 and 
1.25 $\mu$m.\\
(3) An estimate for the contribution by Diffuse Galactic emission is
obtained from recent {\em Spitzer} measurements by \citet{bo}. It is
found that the sum of the  Diffuse Galactic emission and the 
integrated light of resolved galaxies can explain the claimed EBL signal
in the 3.5  $\mu$m window where the ZL contamination is at minimum.\\


\begin{thebibliography}{99}
\bibitem[\protect\citeauthoryear{Aharonian et al.}{2006}]{ah} 
Aharonian F. et al., 2006, Natur, 440, 1018 
\bibitem[\protect\citeauthoryear{Arendt et al.}{1998}]{a1} Arendt, R.G. et al.,
1998 ApJ, 508, 74
\bibitem[\protect\citeauthoryear{Arendt \& Dwek}{2003}]{ar03} Arendt R.~G., Dwek E., 2003, ApJ, 585, 305 
\bibitem[\protect\citeauthoryear{Bernard et al.}{1994}]{be} Bernard, J.P. et al., Boulanger F., Desert F.~X., 
Giard M., Helou G., Puget J.~L., 1994, A\&A, 291, L5  
\bibitem[\protect\citeauthoryear{Bernstein, Freedman \& Madore}{2002a}]{b1} Bernstein, R.A., Freedman, W.L., Madore, B.F., 2002a,
ApJ, 571, 56
\bibitem[\protect\citeauthoryear{Bernstein, Freedman \& Madore}{2002b}]{b2} Bernstein, R.A., Freedman, W.L., Madore, B.F., 2002b,
ApJ, 571, 85
\bibitem[\protect\citeauthoryear{Bernstein, Freedman \& Madore}{2005}]{b5} Bernstein, R.A., Freedman, W.L., Madore, B.F., 2005,
ApJ, 632, 713
\bibitem[\protect\citeauthoryear{Bohlin, Savage, \& Drake}{1978}]{boh} Bohlin R.~C., Savage B.~D., Drake J.~F., 
1978, ApJ, 224, 132 
\bibitem[\protect\citeauthoryear{Boulanger et al.}{1996}]{bou96} Boulanger F., Abergel A., Bernard J.-P., 
Burton W.~B., Desert F.-X., Hartmann D., Lagache G., Puget J.-L., 1996, 
A\&A, 312, 256 
 \bibitem[\protect\citeauthoryear{Cambr\'esy et al.}{2001}]{c1} Cambr\'esy, L., Reach, W.T.,
Beichman, C.A., Jarret. T.H., 2001, ApJ, 555, 563
\bibitem[\protect\citeauthoryear{Colina, Bohlin, R.C.\& Castelli}{1996}]{c2} Colina,L., Bohlin, R.C., Castelli, F.
1996, AJ, 112, 307 
\bibitem[\protect\citeauthoryear{Cooray \& Yoshida}{2004}]{cooray} 
Cooray A., Yoshida N., 2004, MNRAS, 351, L71 
\bibitem[\protect\citeauthoryear{Dwek \& Arendt}{1998}]{dw98} Dwek E., Arendt R.~G., 1998, ApJ, 508, L9 
\bibitem[\protect\citeauthoryear{Arendt \& Dwek}{2003}]{da03} Arendt R.~G., Dwek E., 2003, ApJ, 585, 305 
\bibitem[\protect\citeauthoryear{Dwek, Arendt, \& 
Krennrich}{2005}]{dw05} Dwek E., Arendt R.~G., Krennrich F., 2005, ApJ, 635, 784 
\bibitem[\protect\citeauthoryear{Fazio et al.}{2004}]{fa} Fazio G.~G., et al., 2004, ApJS, 154, 39 
\bibitem[\protect\citeauthoryear{Flagey et al.}{2006}]{bo} Flagey N., Boulanger F., Verstraete L., Miville Desch{\^e}nes M.~A., 
Noriega Crespo A., Reach W.~T., 2006, A\&A, 453, 969
\bibitem[\protect\citeauthoryear{Gorjian, Wright \& Charry}{2000}]{g1} Gorjian, V., Wright, E.L.,
Charry, R.R., 2000, ApJ, 536, 550
\bibitem[\protect\citeauthoryear{Hauser \& Dwek}{2001}]{h01} Hauser M.~G., Dwek E., 2001, ARA\&A, 39, 
249
\bibitem[\protect\citeauthoryear{Kalberla et al.}{2005}]{ka} Kalberla, P. M. W.; Burton, W. B.; 
Hartmann, Dap; Arnal, E. M.; Bajaja, E.; Morras, R.; Pöppel, W. G. L., 2005, A\&A 440, 775 
\bibitem[\protect\citeauthoryear{Kashlinsky et al.}{2005}]{kash} Kashlinsky A., Arendt R.~G., Mather J., 
Moseley S.~H., 2005, Natur, 438, 45 
\bibitem[\protect\citeauthoryear{Kelsall et al.}{1998}]{k1} Kelsall, T. et al. 
1998 ApJ, 508, 44
\bibitem[\protect\citeauthoryear{Laureijs, Mattila \& Schnur}{1987}]{lms} Laureijs R.~J., 
Mattila K., Schnur G., 1987, A\&A, 184, 269 
\bibitem[\protect\citeauthoryear{Lehtinen \& Mattila}{1996}]{lm} Lehtinen K., Mattila K., 1996, 
A\&A, 309, 570 
\bibitem[\protect\citeauthoryear{Leinert et al.}{1998}]{l1} Leinert, C. et al., 
1998, A\&AS 127, 1 
 \bibitem[\protect\citeauthoryear{Longair}{2001}]{lo1} Longair M., 2001, IAUS, 204, 505 
\bibitem[\protect\citeauthoryear{Madau \& Silk}{2005}]{madau} Madau, P., Silk, J., 2005, MNRAS, 359, 37
\bibitem[\protect\citeauthoryear{Madau \& Pozzetti}{2000}]{mapo} Madau P., Pozzetti L., 2000, MNRAS, 312, L9 
\bibitem[\protect\citeauthoryear{Mathis}{1990}]{mathis} Mathis 
J.~S., 1990, ARA\&A, 28, 37 
\bibitem[\protect\citeauthoryear{Mattila}{1979}]{m79} 
Mattila K., 1979, A\&A, 78, 253 
\bibitem[\protect\citeauthoryear{Mattila}{1980}]{m80} Mattila K., 1980, A\&AS, 39, 53 
\bibitem[\protect\citeauthoryear{Mattila}{1990}]{m90} 
Mattila K., 1990, IAUS, 139, 257  
\bibitem[\protect\citeauthoryear{Mattila}{2003}]{m2} Mattila, K., 2003, ApJ, 591, 119
\bibitem[\protect\citeauthoryear{Matsumoto et al.}{2005}]{m1} Matsumoto, T. et al., 2005
ApJ, 626, 31
\bibitem[\protect\citeauthoryear{Salvaterra \& Ferrara}{2003}]{salvaterra} 
Salvaterra R., Ferrara A., 2003, MNRAS, 339, 973
\bibitem[\protect\citeauthoryear{Santos, Bromm, \& Kamionkowski}{2002}]{santos} Santos M.~R., Bromm V., 
Kamionkowski M., 2002, MNRAS, 336, 1082 
\bibitem[\protect\citeauthoryear{Savage \& Oliver}{2005}]{savage} Savage, R. S., Oliver, S., 2005 (astro-ph/0511359)
\bibitem[\protect\citeauthoryear{Sellgren, Werner \& Dinerstein}{1983}]{s1}Sellgren, K., Werner, M. W., 
Dinerstein, H. L., 1983, ApJ, 271, L13
\bibitem[\protect\citeauthoryear{Totani et al.}{2001}]{totani} 
Totani T., Yoshii Y., Iwamuro F., Maihara T., Motohara K., 2001, ApJ, 550, L137 
\bibitem[\protect\citeauthoryear{Wright}{1998}]{w98} Wright 
E.~L., 1998, ApJ, 496, 1 
\bibitem[\protect\citeauthoryear{Wright}{2001}]{w1} Wright, E.L., 2001, ApJ, 553, 538
\bibitem[\protect\citeauthoryear{Wright \& Reese}{2000}]{wr00} Wright E.~L., Reese E.~D., 2000, ApJ, 
545, 43 

\end{thebibliography}
 \end{document}